\documentclass[aps,12pt,final,notitlepage,oneside,onecolumn,nobibnotes,nofootinbib,
superscriptaddress,noshowpacs,centertags]{revtex4-1}
\usepackage[utf8]{inputenc}
\usepackage[english]{babel}
\usepackage{amsmath}
\usepackage{amsfonts}
\usepackage{amssymb}
\usepackage{graphicx}
\newcommand{\cD}{\mathcal{D}}
\newcommand{\cF}{\mathcal{F}}
\newcommand{\cH}{\mathcal{H}}
\newcommand{\cN}{\mathcal{N}}
\newcommand{\R}{\mathbb{R}}
\newcommand{\Ei}{\mathrm{Ei}}
\newcommand{\bra}{\langle}
\newcommand{\ket}{\rangle}
\newcommand{\eg}{{\sl e.g. }}
\def\dk#1#2{\frac{ d^{#2}{#1} }{ (2\pi)^{#2} }} 
\begin{document}
\title{Wavelets and renormalization group in quantum field theory problems}
\author{Altaisky M.V.}
\email{altaisky@rssi.ru}
\affiliation{Space Research Institute RAS, Profsoyuznaya 84/32, Moscow, 117997, Russia}

\begin{abstract}
Using continuous wavelet transform it is possible to construct a regularization 
procedure for scale-dependent quantum field theory models, which is complementary 
to functional renormalization group method in the sense that it sums up the 
fluctuations of larger scales in order to get the effective action at small 
observation scale (M.V.Altaisky, {\em Phys. Rev. D} {\bf 93}(2016) 105043). The standard RG 
results for $\phi^4$ model are reproduced. The fixed points 
of the scale-dependent theory are studied in one loop approximation.
\end{abstract}
\maketitle
\section{Introduction}
Renormalization group (RG) has entered quantum field theory as a group  
of infinitesimal reparametrizations of the $S$ matrix emerging after the cancellation of ultraviolet divegences \cite{SP1953}. The RG method has become known in quantum electrodynamics (QED) since Gell-Mann and Low have shown the charge distribution surrounding a 
test charge in vacuum does not at small distances depend on a coupling constant, except for a scale factor, {\em i.e.,} possesses a kind of self-similarity, that enables to 
express a ''bare'' charge at small scale using the measured value at large scale \cite{GL1954}. 

RG can be considered as a method of treating physical problems with a large number degrees of freedom , not taking all those at once, but treating them   successively scale-by-scale \cite{Wilson1971a,WK1974}. This resulted in an elegant theory of critical phenomena and 
was later generalized to many other stochastic systems \cite{Vasiliev2004}. 

Same idea of separating the fluctuations of different scales has been 
implemented, basically in experimental data processing, in a quite different way: using 
wavelets. This was first done in geophysics \cite{GGM1984,GM1984}, and then spread over 
all possible data, from face recognition and medical imaging to high energy physics and cosmology \cite{Daub10}. The only interference of the RG and the wavelet method seems 
to be the the lattice regularization in quantum field theory (QFT), which can be performed either by standard lattice methods, or by using the discrete wavelet basis \cite{Caroll1993,Battle1999,Best2000}. 
The connections between these two seemingly different methods are still missing. The text below is an endeavor to fill this gap partially.

\section{Divergences in quantum field theory}
The fundamental problem of quantum field theory is the problem of divergences of Feynman graphs. The infinities appearing in perturbation 
expansion of Feynman integrals are treated by different regularization methods, from maximal momentum cutoff and Pauli-Villars regularization, to $\epsilon$-expansion, and renormalization group methods, see \eg \cite{Collins1984} for a review. We restrict ourselves with a 
simple example of scalar $\phi^4$ field theory in $\R^d$, which, however, 
illustrates all main problems and approaches related to the 
problem of divergences in quantum field theory,  see \eg \cite{Collins1984,Ramond1989}. 

Euclidean scalar field theory with $\phi^4$ interaction potential is 
determined by the generating functional 
\begin{align}
Z[J] &=& \cN \int \exp\left(-S_E[\phi] + \int J(x)\phi(x) d^dx\right) \equiv \exp(W[J]) \label{gf}\\
\nonumber \hbox{where\ } & &S_E[\phi] = \int d^dx \left[ \frac{1}{2} (\partial\phi)^2 + \frac{m^2}{2} \phi^2  
+ \frac{\lambda}{4!}\phi^4\right]. 
\end{align}
where $\cN$ is a formal normalization constant.
The connected Green functions, understood as statistical momenta of 
the field $\phi$ \cite{ZJ1999}, are given 
by functional derivatives of the generating functional:
\begin{equation}
G^{(n)} \equiv \bra\phi(x_1)\ldots\phi(x_n) \ket_c = 
\left. { \frac{\delta^n\ln W[J]}{\delta J(x_1) \ldots \delta J(x_n)}
}\right|_{J=0}
\label{cgf}
\end{equation}
The divergences of Feynman graphs in the perturbation expansion 
of the Green functions \eqref{cgf} with respect to the small 
coupling constant $\lambda$  emerge at coinciding arguments 
$x_i=x_k$. For instance, the bare two-point Green  
function 
\begin{equation}
G^{(2)}_0(x-y) = \int \frac{d^dp}{(2\pi)^d}\frac{e^{-\imath p(x-y)}}{p^2+m^2}
\end{equation}
is divergent at $x\!=\!y$ for $d\ge2$.

Since the Green functions in Euclidean quantum field theory 
have the probability meaning, it is quite obvious 
physically
that neither of the joint probabilities of the measured quantities can be 
infinite. The infinities seem to be caused by an  
inadequate choice of the functional space the fields belong to.

This standard approach to quantum field theory, based on $\mathrm{L}^2(\R^d)$ fields disregards two important notes \cite{Altaisky2010PRD}:
\begin{enumerate}
\item To localize a particle in an interval $\Delta x$ the measuring device 
requests a momentum transfer of order $\Delta p\!\sim\!\hbar/\Delta x$. If 
the value of this momentum is too large we may get out of the applicability 
range of the initial model, in the sense that $\phi(x)$ at 
a fixed point $x$ has no experimentally verifiable meaning. What is meaningful, is the vacuum expectation of product of fields in certain 
region centered around $x$, the width of which ($\Delta x$) is constrained 
by the experimental conditions of the measurement.  
\item Even if the particle has been 
initially prepared on the interval 
$(x-\frac{\Delta x}{2},x+\frac{\Delta x}{2})$, the probability of 
registering it on this interval is generally less than unity:
for the probability of registration depends on the strength of interaction 
and the ratio of typical scales of the measured particle and the measuring 
equipment. The maximum probability of registering an object of 
typical scale $\Delta x$ by the equipment with typical resolution $a$
 is achieved when these two parameters are comparable. For this reason 
the probability of registering an electron by visual range photon scattering 
is much higher than by that of long radio-frequency waves. As 
mathematical generalization, we should say that if a measuring equipment  
with a given spatial resolution $a$ fails to register an object, prepared 
on spatial interval of width $\Delta x$ with certainty, 
then tuning the equipment to {\em all} possible resolutions $a'$ would 
lead to the registration. This certifies the fact of the existence 
of the object.    
\end{enumerate}

Most of the regularization methods applied to make the Green functions 
finite imply a certain type of self-similarity -- the independence of physical 
observables on the scale transformation of an arbitrary parameter of the theory --  
the cutoff length or the normalization scale. Covariance with respect 
to scale transformations is expressed by renormalization group equation 
\cite{Collins1984}. Its predecessor, the Kadanoff blocking 
procedure, averages the small-scale fluctuations up to a certain scale 
into a kind of effective interaction for a larger blocks, assuming the larger 
blocks interact with each other in the same way as their sub-blocks  
\cite{Kadanoff1966,Ito1985}. 
However the theory based on the Fourier 
transform, such as quantum field theory is, have no explicit tools to 
regard the self-similarity in a fair way.  
 An abstract harmonic analysis based on some group $G$, wider than 
the group of translations $G:x\to x+b$, should be used to account for self-similarity. The simplest analysis of such type is based on the representations of affine group $G:x \to ax+b$ and is widely referred to 
as continuous wavelet transform.

\section{Continuous wavelet transform in quantum field theory}

Continuous wavelet transform (CWT) is a  generalization of the Fourier transform 
for the case when the scaling properties of the theory are important. 
Referring the reader to general reviews on wavelet transform \cite{Daub10,Chui1992}, and 
to the original papers devoted to the application of wavelet transform to quantum 
field theory \cite{Altaisky2010PRD,AK2013,BP2013}, below we remind basic definitions 
of the wavelet formalism of quantum field theory. 

Let $\cH$ be a Hilbert space of states for a quantum field $|\phi\ket$. 
Let $G$ be a locally compact Lie group acting transitively on $\cH$, 
with $d\mu(\nu),\nu\in G$ being a left-invariant measure on $G$.  
Similarly to the Fourier representation 
$
|\phi\ket=\int |p\ket dp \bra p |\phi\ket,$
any $|\phi\ket \in \cH$ can be decomposed with respect to 
a representation $U(\nu)$ of $G$ in $\cH$ \cite{Carey1976,DM1976}:
\begin{equation}
|\phi\ket= \frac{1}{C_g}\int_G U(\nu)|g\ket d\mu(\nu)\bra g|U^*(\nu)|\phi\ket, \label{gwl} 
\end{equation} 
where $|g\ket \in \cH$ is referred to as an admissible vector, 
or a {\em basic wavelet}, satisfying the admissibility condition 
$
C_g = \frac{1}{\| g \|^2} \int_G |\bra g| U(\nu)|g \ket |^2 
d\mu(\nu)
<\infty. 
$
The coefficients $\bra g|U^*(\nu)|\phi\ket$ are referred to as 
wavelet coefficients. 
If the group $G$ is Abelian, the wavelet transform \eqref{gwl} coincides with 
the Fourier transform. 

Next to the Abelian group is the group of the affine transformations 
of the Euclidean space $\R^d$:
\begin{equation}
G: x' = a x + b, x,b \in \R^d, a \in \R_+.  \label{ag1}
\end{equation} 
(For simplicity we assume the isotropic basic wavelet $g$ and 
drop rotation factor.)
The unitary representation of the affine transform \eqref{ag1} with 
respect to the isotropic basic wavelet $g(x)$ can be written as follows:
\begin{equation}
U(a,b) g(x) = \frac{1}{a^d} g \left(\frac{x-b}{a} \right).
\end{equation}  
(In accordance to previous papers \cite{Altaisky2010PRD,AK2013} we use $L^1$ norm \cite{Chui1992,HM1996}  to keep the physical dimension 
of wavelet coefficients equal to the dimension of the original fields).

Wavelet coefficients of the Euclidean field $\phi(x)$ with 
respect to the basic wavelet $g(x)$ in  $\R^d$ are  
\begin{equation}
\phi_{a}(b) = \int_{\R^d} \frac{1}{a^d} \overline{g \left(\frac{x-b}{a} \right) }\phi(x) d^dx. \label{dwtrd}
\end{equation} 

The function $\phi(x)$ can be reconstructed from its wavelet coefficients 
\eqref{dwtrd} using the formula \eqref{gwl}:
\begin{equation}
\phi(x) = \frac{1}{C_g} \int \frac{1}{a^d} g\left(\frac{x-b}{a}\right) \phi_{a}(b) \frac{dad^db}{a}.  \label{iwt}
\end{equation}
The normalization 
constant is readily evaluated using Fourier transform:
$
C_g = \int_0^\infty |\tilde g(a)|^2\frac{da}{a}.
$

Substituting \eqref{iwt} into the field theory \eqref{gf} we obtain the 
generating functional for the scale-dependent fields $\phi_a(x)$: 
\begin{align} \nonumber 
Z_W[J_a] &=&\cN \int \cD\phi_a(x) \exp \Bigl[ -\frac{1}{2}\int \phi_{a_1}(x_1) D(a_1,a_2,x_1-x_2) \phi_{a_2}(x_2)
\frac{da_1d^dx_1}{a_1}\\
\nonumber &\times& \frac{da_2d^dx_2}{a_2}  
-
\int V_{x_1,\ldots,x_4}^{a_1,\ldots,a_4} \phi_{a_1}(x_1)\cdots\phi_{a_4}(x_4)
\frac{da_1 d^dx_1}{a_1} \frac{da_2 d^dx_2}{a_2} \\
&\times& \frac{da_3 d^dx_3}{a_3} \frac{da_4 d^dx_4}{a_4} 
+ \int J_a(x)\phi_a(x)\frac{dad^dx}{a}\Bigr], \label{gfw}
\end{align}
with $D(a_1,a_2,x_1-x_2)$ and $V_{x_1,\ldots,x_4}^{a_1,\ldots,a_4}$ denoting the wavelet images of the inverse propagator and that of the interaction potential.

The Feynman diagram technique for the scale-dependent fields $\phi_a(x)$ is the same 
as for ordinary fields except for \cite{Alt2002G24J,Altaisky2010PRD}:
\begin{itemize}\itemsep=0pt
\item each field $\tilde\phi(k)$ will be substituted by the scale component: 
$\tilde\phi(k)\to\tilde\phi_a(k) = \overline{\tilde g(ak)}\tilde\phi(k)$.
\item each integration in momentum variable is accompanied by corresponding 
scale integration:
\[
 \dk{k}{d} \to  \dk{k}{d} \frac{da}{a}.
 \]
\item each interaction vertex is substituted by its wavelet transform; 
for the $N$-th power interaction vertex this gives multiplication 
by factor 
$\displaystyle{\prod_{i=1}^N \overline{\tilde g(a_ik_i)}}$.
\end{itemize}
According to these rules, the bare Green function in wavelet representation 
takes the form 
$$
G^{(2)}_0(a_1,a_2,p) = \frac{\tilde g(a_1p)\tilde g(-a_2p)}{p^2+m^2}.
$$ 
The finiteness of the loop integrals is provided by the following rule:
{\em there should be no scales $a_i$ in internal lines smaller than the minimal scale 
of all external lines} \cite{Altaisky2010PRD}. Therefore the integration in $a_i$ variables is performed from 
the minimal scale of all external lines up to the infinity. 
This corresponds to the assumption, that studying a system from outside one should 
not used functions with resolution better than the finest experimentally available 
scale. 
The integration over {\sl all} scales will certainly drive us back to the known divergent theory. 

\section{Renormalization group for scale-dependent fields}
The only intersection between usual regularization methods of quantum field 
theory and the separation of different scales by means of wavelet transform 
up to very recently was the lattice regularization in wavelet basis \cite{Caroll1993,Battle1999,Best2000,BP2013}. In \cite{Alt2016prd} it 
was proposed to use CWT for regularization of 
quantum field theory along the lines the renormalization group is usually applied. To illustrate this let us rewrite the formalism of functional 
renormalization group for the effective averaging action, which 
accounts for the fluctuations with momentum $k$ integrating over fluctuations with momenta greater than $k$.

Let  $\cF_k$ be the space of functions the Fourier images of which are 
supported by $|p|\le k$ domain. The effective action $S_k[\phi]$ is defined 
via 
$$
e^{-S_k[\phi]}= \int \cD \chi P_k[\phi,\chi] e^{-S[\chi]},
$$
where $P_k[\phi,\chi]$ is a projection of $\chi$ onto the space $\cF_k$ \cite{Wetterich1991}. If we know the action $S_k[\phi]$ we can coarse-grain 
to the next space $\cF_{k-\Delta k}$ integrating over the functions $\tilde{\phi}\in \cD_{k,\Delta k}= \cF_{k}\setminus\cF_{k-\Delta k}$, whose 
momenta are within the range $(k-\Delta k,k]$. The iteration of this 
procedure 
\begin{equation}
e^{-S_{k-\Delta k}[\phi]} = \int \cD[\tilde{\phi}] e^{-S_k[\phi+\tilde{\phi}]}, \label{frg}
\end{equation}
back to arbitrary small (IR) $k$ yields the scale 
decomposition 
\begin{align} \nonumber  
\ldots \subset \cF_{k-2\Delta k} \subset \cF_{k-\Delta k} \subset \cF_k, \\
\cF_k = \cD_{k,\Delta_k} \oplus \cD_{k-\Delta k,\Delta k} \oplus \cD_{k-2\Delta k,\Delta k} \oplus \ldots ,
\end{align}
very similar to the Mallat sequence in wavelet analysis \cite{Ma1986}. Doing so, 
one get an exact action $\Gamma[\phi]= \lim_{k\to0}S_k[\phi]$. 
This is the essence of functional RG.  

Working with wavelet transform we can do something complementary to 
functional renormalization group: we can sum up all fluctuations from 
infinitely large IR scale to a certain finite scale of observation $A$ to 
obtain an effective action functional, which describes the physics 
at scale $A$. To work with 1PI diagrams we define the effective 
action via the Legendre transform of $W[J]$:
\begin{equation}
\Gamma[\phi_a] =   -W_W[J_a] + \int J_a(x) \phi_a(x) \frac{da}{a}d^dx, \label{Gw}
\end{equation}
The functional derivatives of $\Gamma[\phi]$ are the renormalized vertex 
functions $\Gamma^{(n)}$. 

\begin{figure}[ht]
\centering \includegraphics[width=60mm]{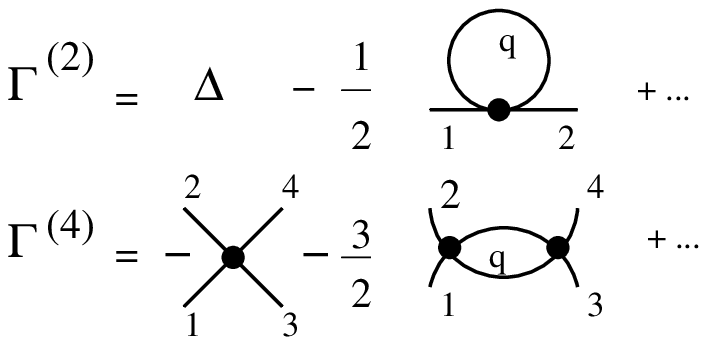}
\caption{Renormalized inverse propagator and the renormalized vertex functions in $\phi^4$ theory shown in one-loop approximation. Here and after we assume $-\lambda$ value for each vertex. Redrawn from \cite{Alt2016prd}}
\label{v24:pic}
\end{figure}

Following \cite{Alt2016prd} we consider $\Gamma^{(2)}$ and $\Gamma^{(4)}$ 
vertex functions for $\phi^4$ theory in $\R^d$ in one-loop level. The one 
loop contributions to the inverse propagator $\Gamma^{(2)}_{(A)}$ and the 
vertex function $\Gamma^{(4)}_{(A)}$ are given by  diagrams 
shown in Fig.~\ref{v24:pic} 
After integration in scale arguments $d \ln a_i$ of the internal lines, 
the only difference between the wavelet-based theory and the standard one will be the presence of the squared cutoff functions $f^2(x)$
 on each 
internal line, depending on the dimensionless momenta of the line $x=qA$. 
This gives 
\begin{equation}
C_g^2\frac{\Gamma^{(2)}_{(A)}(a_1,a_2,p)}{\tilde{g}(a_1 p) \tilde{g}(-a_2 p)} = p^2+m^2 + 
\frac{\lambda}{2} T^d_g(\alpha) \label{g2l1},
\end{equation} 
where $d=4$ is the dimension of Euclidean space,
$\alpha = m \min (a_1,a_2)$ is the dimensionless scale of the tadpole diagram, for the inverse 
propagator; and similarly, 
\begin{equation}
C_g^4 \frac{\Gamma^{(4)}_{(A)}}{\tilde{g}(a_1p_1)\tilde{g}(a_2p_2)\tilde{g}(a_3p_3)\tilde{g}(a_4p_4) } = \lambda -\frac{3}{2}\lambda^2 X^d_g(A) \label{g4l1}
\end{equation}
for the vertex function. 

The values of the one-loop integrals 
\begin{align}\nonumber 
T^d_g(\alpha) &=& \frac{S_d m^{d-2}}{(2\pi)^d} \int_0^\infty f_g^2(\alpha x) \frac{x^{d-1}dx}{x^2+1}, \\\ 
X^d_g(A) &=& \int \frac{d^dq}{(2\pi)^d}
\frac{f^2_g(qA)f^2_g((q-s)A)}{\left[ q^2+m^2\right]\left[ (q-s)^2+m^2\right] },
\label{li1}
\end{align}
where $s\!=\!p_1\!+\!p_2, A=\min(a_1,a_2,a_3,a_4)$, 
depend on the wavelet cutoff function 
\begin{equation}
f_g(x) = \frac{1}{C_g}\int_x^\infty |\tilde{g}(a)|^2 \frac{da}{a}
\end{equation}
for the chosen wavelet $g$. 

The dependence of the effective coupling constant on the observation scale 
$A$ can be obtained by taking the derivative with respect to the logarithm 
of observation scale $\mu = -\ln A + const$. For the $g_1$ wavelet in $d=4$ used in \cite{Alt2016prd} this gives the flow equations 
 \begin{align}
\frac{\partial \lambda}{\partial\mu} &=& 3\lambda^2\alpha^2 \frac{\partial X^4_1}{\partial\alpha^2} = \frac{3\lambda^2}{16\pi^2} \frac{2\alpha^2+1-e^{\alpha^2}}{\alpha^2}
e^{-2\alpha^2}, \label{b1} \\
\frac{1}{m^2} \frac{\partial m^2}{\partial\mu} &=& \frac{\lambda}{32\pi^2\alpha^2}
-\frac{\lambda}{16\pi^2} + \frac{\lambda}{16\pi^2} 2\alpha^2e^{2\alpha^2}\Ei_1(2\alpha^2),
\label{b2}
\end{align}
where $\alpha = A m$ is the dimensionless scale,  
$\Ei_1(z)=\int_1^\infty \frac{e^{-xz}}{x}dx$ is the exponential integral of the first type. 
To find the scale dependence of the coupling constant $\lambda=\lambda(\mu)$ explicitly. To do this 
we rewrite \eqref{b1} as 
\begin{equation}
\frac{\partial \lambda}{\partial \mu} 
= \frac{3\lambda^2}{16\pi^2}B(\alpha^2), \quad 
B(x) =  
\left(2 + \frac{1}{x} \right) e^{-2x} - \frac{e^{-x}}{x}
\label{b1a}
\end{equation}
where $x=\alpha^2$ and $d\mu = -\frac{d\alpha}{\alpha}$. The equation above can be solved now for the inverse coupling constant 
$g = \frac{1}{\lambda}$:
\begin{align*}
dg = \frac{3}{32\pi^2} \left\{
\left(2+\frac{1}{x}\right) e^{-2x} - \frac{e^{-x}}{x}
\right\} \frac{dx}{x} \equiv \frac{3}{32\pi^2} dF(x),
\quad 
F(x) = \frac{e^{-x}}{x} - \frac{e^{-2x}}{x} - \mathrm{Ei}_1(x).
\end{align*} 
Inverting the above equation we get the scale dependence of the 
coupling constant  
\begin{equation}
\lambda(x) = \lambda(x; x_1,\lambda_1) = 
\frac{1}{\frac{1}{\lambda_1} + \frac{3}{32\pi^2}
[F(x)-F(x_1)]},
\label{lx}
\end{equation}
where $\lambda_1 \equiv \lambda(x_1)$ is the boundary condition for the coupling constant $\lambda$. The graphs of the ultraviolet behavior of 
$\lambda(x)$ for different infrared boundary conditions $\lambda_1$ are 
shown in Fig.~\ref{l:pic} below.
\begin{figure}[ht]
\centering \includegraphics[width=50mm]{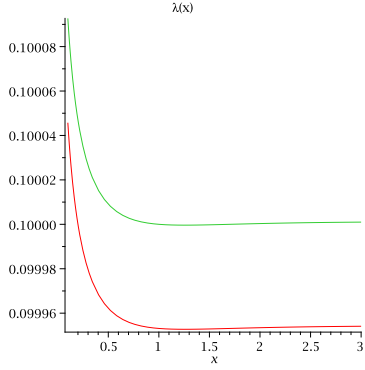}
\caption{Dependence of the coupling constant $\lambda=\lambda(x)$ on the dimensionless scale $x=(Am)^2$}
\label{l:pic}
\end{figure}

\begin{figure}[ht]
\centering \includegraphics[width=50mm]{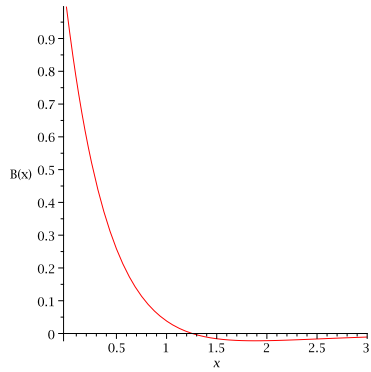}
\caption{Dependence of the beta function $B(x)$ on the dimensionless scale $x=(Am)^2$}
\label{Bx:pic}
\end{figure}

The zeros of the $\beta$-function $\beta(\lambda,\mu) = \frac{3\lambda^2}{16\pi^2} B(x)$ 
except for the trivial case $\lambda=0$ are determined by the equation  
\begin{equation}
B(x) = \left(2+\frac{1}{x}\right) e^{-2x} - \frac{e^{-x}}{x}=0
\end{equation}
The graph of the function $B(x)$ is shown in Fig.~\ref{Bx:pic}

The solutions of the equation $B(x)=0$ are given 
by the equality 
$
2x+1 = e^{x},
$
which can be satisfied for either $x=0$ or 
$$
x_* = -\mathrm{LambertW}(-1,-\frac{1}{2}e^{-1/2}) - \frac{1}{2}\approx 
1.25643 \ldots
$$

Let there exist a fixed point value of the coupling constant 
$\lambda_* = \lambda(x_*)$, then, as it follows from the 
graph shown above:
\begin{itemize}
\item if $\lambda > \lambda_*$ to the left from $x_*$, the decrease of $\mu \to 0$ results in the decrease of $\lambda$
\item if $\lambda < \lambda_*$ to the right from $x_*$, the increase of $\mu$ results in increase of $\lambda$
\end{itemize}
Therefore $\lambda_*$ is an IR stable fixed point.
%

\end{document}